\documentclass[12pt]{iopart}
\usepackage{graphics,graphicx}
\begin{document}

\title[Melting of colloidal molecular crystals] {Phase behaviour of
  colloidal assemblies on 2D corrugated substrates}

\author{Samir El Shawish$^1$, Emmanuel Trizac$^2$ and Jure
  Dobnikar$^{3,4}$} 

\address{$^1$Reactor Engineering Division, Jo\v zef Stefan Institute,
  Jamova 39, 1000 Ljubljana, Slovenia \\ $^2$Universit\'e Paris-Sud,
  Laboratoire de Physique Th\'eorique et Mod\`eles Statistiques (CNRS
  UMR 8626), 91405 Orsay Cedex (France) \\ $^3$ Department of
  Theoretical Physics, Jo\v zef Stefan Institute, Jamova 39, 1000
  Ljubljana, Slovenia\\ $^4$ Department of Chemistry, University of
  Cambridge, Lensfield Road, CB2 1EW, Cambridge, UK }
\eads{\mailto{jd489@cam.ac.uk},\mailto{trizac@lptms.u-psud.fr}}

\begin{abstract}
  We investigate - with Monte Carlo computer simulations - the phase
  behaviour of dimeric colloidal molecules on periodic substrates with
  square symmetry. The molecules are formed in a two-dimensional
  suspension of like charged colloids subject to periodic external
  confinement, which can be experimentally realized by optical
  methods. We study the evolution of positional and orientational
  order by varying the temperature across the melting transition. We
  propose and evaluate appropriate order parameters as well as the
  specific heat capacity and show that the decay of positional
  correlations belongs to a class of crossover transitions while the
  orientational melting is a second order phase transition.
\end{abstract}
\submitto{\JPCM}

\section{Introduction}
\label{sec:intro}

Understanding, creating and controlling patterned structures on nano-
and microscale is an important objective of modern materials science
\cite{nanopattern}. Studying colloidal ordering at surfaces provides
valuable insights into the underlying mechanisms. Complex surfaces
with submicron periodicities can direct self-assembly of highly
oriented responsive colloidal crystals \cite{Yodh2000} and
biologically active substrates \cite{dogic2010,biosensors}. Charged
objects are ubiquitous in colloidal domain and electrostatic
interactions have been widely employed in colloidal
assembly. Predominantly spherical colloidal particles have been
studied, while comparatively little work has been devoted to the
behaviour of anisotropic charged composite objects in a solution. The
spherical shape is, however, more an exception than a rule in the
colloidal realm, and from the fundamental point of view, understanding
the mechanisms directing the assembly of anisotropic charged colloids
is at least equally important. Moreover, it enables design of novel
ordered micro- and nanostructures.

Experimentally, patterned substrates are typically created by either
optical manipulation \cite{VanBlaaderen2002} or surface deposition
\cite{template,deposit}. Multiple charged spherical colloids can then
be confined within the traps and if the number of colloids $N$ is an
integer of the number of traps $N_{\rm tr}$, at strong enough
confinement, monodisperse clusters or ``colloidal molecules'' (in
general anisotropic, at least for small $n$ ) constituted of
$n=N/N_{\rm tr}$ colloids are confined to each of the traps. The
coupling between the anisotropy of the charged ``molecules'' and
screening by microions in the solution is surprisingly nontrivial
\cite{Agra,ElShawish,ElShawish2}. The multipole moments generated by
the molecular anisotropy are all screened in a similar fashion, so
that inter-molecular interactions feature non trivial angular
dependence affecting the symmetry of the ordering in a subtle
way. Therefore, at large enough electrostatic interactions,
orientationally ordered ``colloidal molecular crystals'' are
formed. Crystals with remarkably rich variety of orientational
ordering have been observed
\cite{Bech2001,Bech2002,RO2002,RO2005,Agra,Sarlah,ElShawish,ElShawish2,Olson1,ElShawish3}.
In the case of like-charged repulsive colloids the molecular size is
determined by the interplay between the electrostatic repulsion and
the confinement forces, while the intermolecular electrostatic
interactions determine the type of rotational ordering of the
molecules. The ground state behaviour of such colloidal molecular
crystals has been explored in several studies
\cite{Agra,Sarlah,ElShawish,ElShawish2,ElShawish3}, however, in order
to assess their thermal stability the complete phase behaviour needs
to be explored. In \cite{RO2002} molecular dynamics simulations have
been performed showing the generic phase behaviour of colloidal
molecular crystals ($n=2, 3$ and 4) on square and triangular periodic
substrates. A two-stage melting has been reported where first the
orientational and then the positional order vanishes. Similarly, in an
experiment \cite{Bech2001,Bech2002} laser induced freezing and melting
has been observed: upon increasing the confinement strength colloidal
trimers confined to triangular lattice went through a transition from
liquid to crystal and back to modulated liquid. Here, we focused on
colloidal dimers ($n=2$) on two-dimensional square periodic substrates
and performed Parallel Tempering Monte Carlo simulations to study the
melting process in detail. We defined appropriate order parameters
describing the two-stage melting and characterized the order-disorder
transition for both, the orientational and positional correlations. It
turns out that the orientational melting is a second order phase
transition with diverging specific heat at the transition temperature,
while the decay of the positional order is not a phase transition but
rather a gradual crossover-type transition.

\section{Model and methods}
\label{sec:model}

\begin{figure}[ht]
    \includegraphics[width=0.25\textwidth]{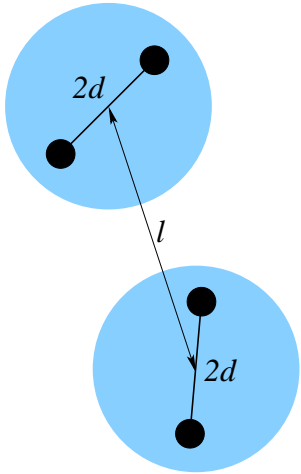}
    \caption{Schematic representation of colloidal dimers in potential
      traps. The distance between the centers of the potential traps
      $l$ is the unit of lenght. In strong enough confinement, exactly
      two colloids occupy each trap forming a colloidal dimer with a
      characteristic size $2d$. The colloidal ``molecules'' are
      well defined, when $2d\ll l$.}
    \label{fig0}
\end{figure}
We consider a system of like charged spherical colloids confined to a
plane and immersed in a suspension of monovalent counterions and salt
ions (the solvent is hereafter considered as a structure-less medium
of constant dielectric permittivity). The microions neutralize the
total charge and mediate repulsive screened Coulomb interactions among
the colloids. We assume the pairwise additive Yukawa potentials and do
not consider many-body contributions to the electrostatic interactions
because the geometry of the problem and the typical screening
parameters explored are in the regime where pairwise additivity is an
excellent assumption \cite{NJP}. The colloids are subject to
additional periodic external confinement, introduced by laser tweezers
\cite{Bech2001,RO2002} creating a set of potential traps for the
colloids. The number of traps $N_{\rm tr}=N/2$ introduced by the
confinement is exactly half of the number of colloids. The total
interaction energy in the dimensionless form is:
\begin{eqnarray} 
E=\sum_{i=1}^N \sum_{j\ne i}^N
\frac{e^{-\kappa r_{ij}}}{r_{ij}/l} - A\sum_{i=1}^N \cos(2\pi
x_i/l)+\cos(2\pi y_i/l)\;,
\label{eq1} 
\end{eqnarray}
where $(x_i,y_i)$ are the Cartesian coordinates of colloid $i$ in the
plane, $r_{ij}$ the distance between colloids $i$ and $j$, $1/\kappa$
is the Debye screening length. The unit of length $l$ is the inter
trap distance (see Fig. \ref{fig0}), and the parameter $A$ is a measure for
the relative strength of the confinement over the electrostatic
forces. The physical parameters governing the phase behaviour of the
system are $\kappa l$, $A$ and the temperature $T$. At low
temperatures and strong confinement each trap is occupied by exactly 2
colloids forming an elongated colloidal dimer. Due to the dimer-dimer
electrostatic interactions, the dimers order into orientationally
ordered colloidal molecular crystals. Upon raising the temperature
first the orientational order of the dimers decays and eventually it
becomes possible for single colloids to hop between the traps thus
destroying the dimers and forming a modulated liquid state.

\begin{figure}[ht]
    \includegraphics[width=0.75\textwidth]{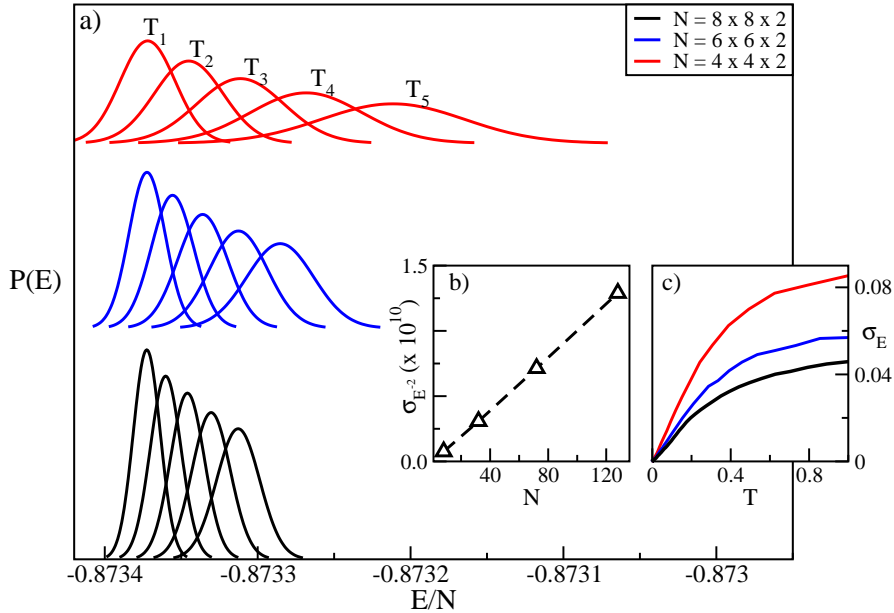}
    \caption{{\bf a)} Energy histograms (Gaussian distributions)
      $P(E)$ calculated at five lowest temperatures $T_1 ... T_5$ in
      equilibrated PTMC simulations for three system sizes $N=32, 72,
      128$ (shifted for clarity). The values of the physical
      parameters are $\kappa l=10$ and $A=0.6$. {\bf Insets:} Standard
      deviation $\sigma_E$ of the distributions $P(E)$ as a function
      of {\bf b)} the number of colloids $N$ at the lowest temperature
      $T=T_1$ and {\bf c)} the temperature $T$ at the values of $N$
      displayed in the main plot.}
    \label{fig1}
\end{figure}

\begin{figure}
\includegraphics[width=0.75\textwidth]{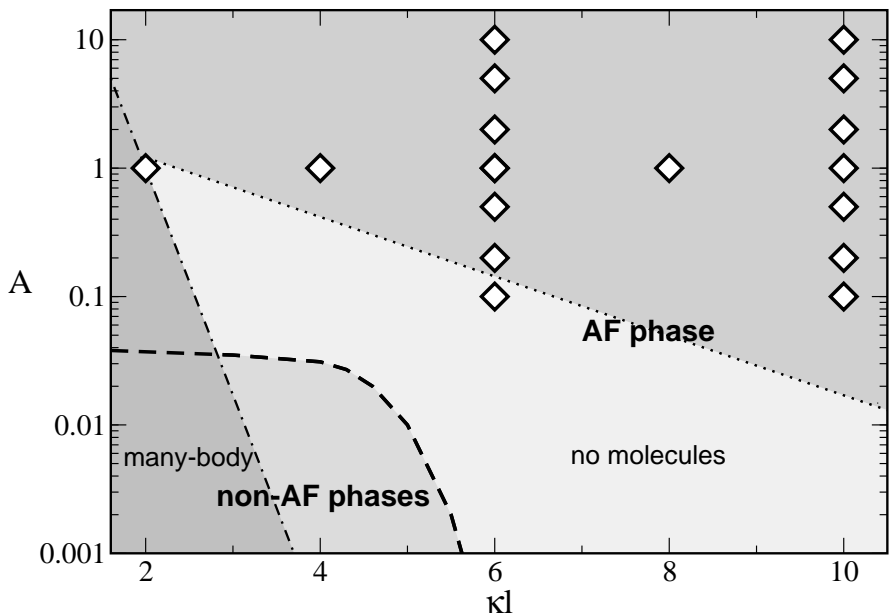}
\caption{T=0 phase diagram of dimers on a square lattice of traps.
  Line separating the AF region from the non-AF region is schematic
  and was calculated on a $N=32$ system using PTMC method.  The
  diamond symbols are placed at the positions in the parameter space
  where the finite temperature behaviour has been explored. The
  regions of the parameter space where many-body interactions are
  dominant and where the molecules are not well defined are marked
  with the shades of grey. Note that phase diagram where cosine
  confinement is replaced by the parabolic trapping \cite{ElShawish}
  is quite richer.}
\label{fig2}
\end{figure}

We used parallel tempering Monte Carlo (PTMC) simulations to simulate
$M$ replicas of the system, each in the canonical ensemble, at
temperatures $T_i$ ranging from $T_1=10^{-4}$ to $T_M=1$ allowing for
configurational swaps between systems with adjacent temperatures
\cite{PTMC} (we chose a geometric temperature profile, thus
$T_i/T_{i+1}={\rm const}$).  Apart from single colloid moves and
configurational swaps we also implemented dimer rotations that become
important at low temperatures. The colloids are treated as point-like
charges and the periodic boundary conditions are applied to the
simulation box. In Fig.\ref{fig1} we show the energy histograms $P(E)$
for our model system at lowest five temperatures and three system
sizes. Overlap of $P(E)$ between adjacent replicas at different
temperatures allows for acceptance of the configuration swaps; in our
simulations the target replica exchange probability is set to
$0.2$. At a given number of colloids $N$, the number of replicas is
chosen accordingly. We found an empiric relation
$M\simeq\sqrt{50 N}$ and studied systems with $(N,M)=(8,20), (32,40),
(72,60), (128,80),$ and (200,100).  From the upper inset in
Fig.\ref{fig1} we see that the width of $P(E)$ (that is proportional
to the standard deviation $\sigma_E$ of the distribution) decreases as
$1/\sqrt{N}$ so that number of replicas $M$ should indeed scale as
$\sqrt{N}$ to maintain a constant overlap. We note that $T_i$ (except
$T_1$ and $T_M$) have different values for different system sizes $N$.
We first examined the ground state phase diagram of dimers ($n=2$) on
a square lattice by PTMC simulations. The parameter space $(\kappa
l,A)$ we have explored is depicted on Fig.\ref{fig2} where several
regions are highlighted. In most of the parameter space on the figure
(right hand side) the ground state configuration of the colloids is
antiferro-like (AF), meaning that the dimers directions in neighbouring
traps are perpendicular.
Structures different from AF are observed
at low $A$ and $\kappa l$: typically, at vanishing confinement
strength $A$, the colloids arrange into a hexagonal lattice and upon
increasing $A$, the structure evolves via I-phases (checkerboard
arrangement with neighbouring dimers' relative angle less than
$\pi/2$, see \cite{ElShawish, Agra}) into AF arrangement (where the
angle between neighbouring dimers is $\pi/2$). At low values of the
screening parameter $\kappa l$, the electric double layers are of
comparable size to the lattice spacing and many-body effects among
colloids are likely to play an important role in orientational
ordering. We have roughly estimated and highlighted the many-body
dominated region with dark grey shading in Fig.\ref{fig2}. Most of the
non-AF phases we discussed above (including those reported in
literature \cite{RO2002} where the parameters were $\kappa l=2$ and
$0<A<0.64$) are in or close to this region, which means that their
stability would need to be re-assessed by applying the many-body
Poisson-Boltzmann theory. The behaviour in this regime is definitely
interesting, however it is experimentally difficult to access and was
not the focus of our present study. We would also like to stress that
in a large part of the parameter space where the confinement is
relatively weak the size of the ``molecules'' in the traps is
significant with respect to the lattice spacing and the notion of
molecule becomes obscure. We have provisionally marked this region
with a light grey shading (please note that there is a certain degree
of arbitrariness in defining the boundaries of such regions).  To
study the phase behaviour, we concentrated on the remaining part of
the parameter space, where molecules are well defined and many-body
effects are negligible. Such regime is also where most of the
experimental studies \cite{Bech2001,Bech2002} have been performed. In
this parameter regime the ground state at $T=0$ is AF phase. We have
analysed the phase behaviour as a function of temperature at the
points described by three sets of parameters (depicted by the diamond
symbols in Fig\ref{fig2}): $0.1\le A\le 10$; $\kappa l=10, 6$ and
$A=1$; $2\le\kappa l\le 10$.

\section{Results}
\label{sec:results}

\begin{figure}[h]
\includegraphics[width=0.75\textwidth]{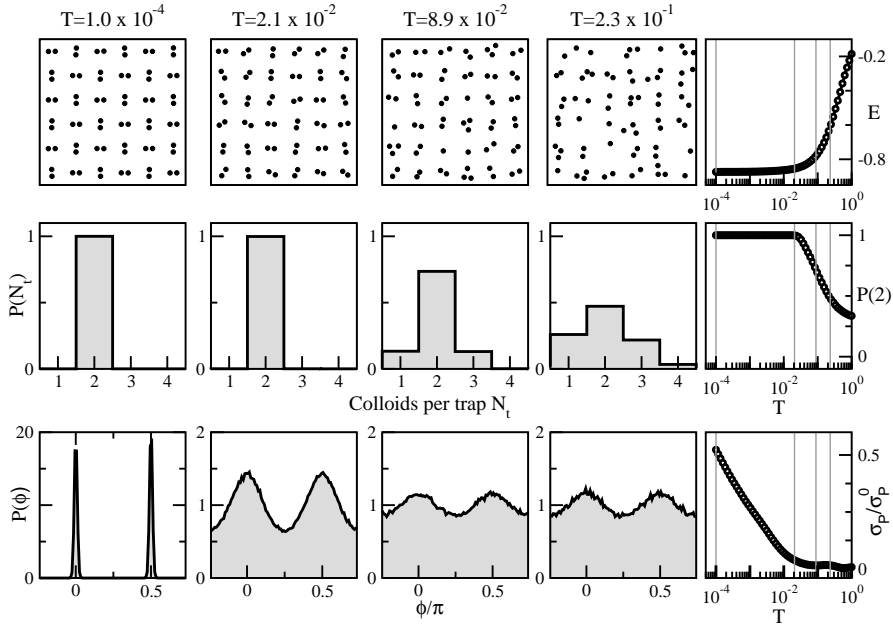}
\caption{Results of the PTMC simulations of 72 colloids subject to
  periodic array of 36 traps at $\kappa l=10$ and $A=0.6$. {\bf Top
    row:} Typical colloidal configurations from the simulations at
  four selected temperatures. The right column shows the energy $E(T)$
  versus temperature $T$ (the four vertical lines denote the temperatures
  of the configurations on left). {\bf Mid row:} Distribution of the
  number of colloids per traps for the same parameters as in the top
  row.  The right column shows the proportion of double-occupied traps
  $P(2)$ as a function of temperature $T$. {\bf Bottom row:} Angle
  distributions $P(\phi)$ of dimers (or higher $n$-mers if there are
  more than two colloids per trap \cite{inertia}) calculated at the
  same four temperatures as above. The right column shows the
  temperature dependence of the spreading parameter as a measure of
  angular anisotropy.}
\label{fig3}
\end{figure}

\begin{figure}[ht]
\includegraphics[width=0.8\textwidth]{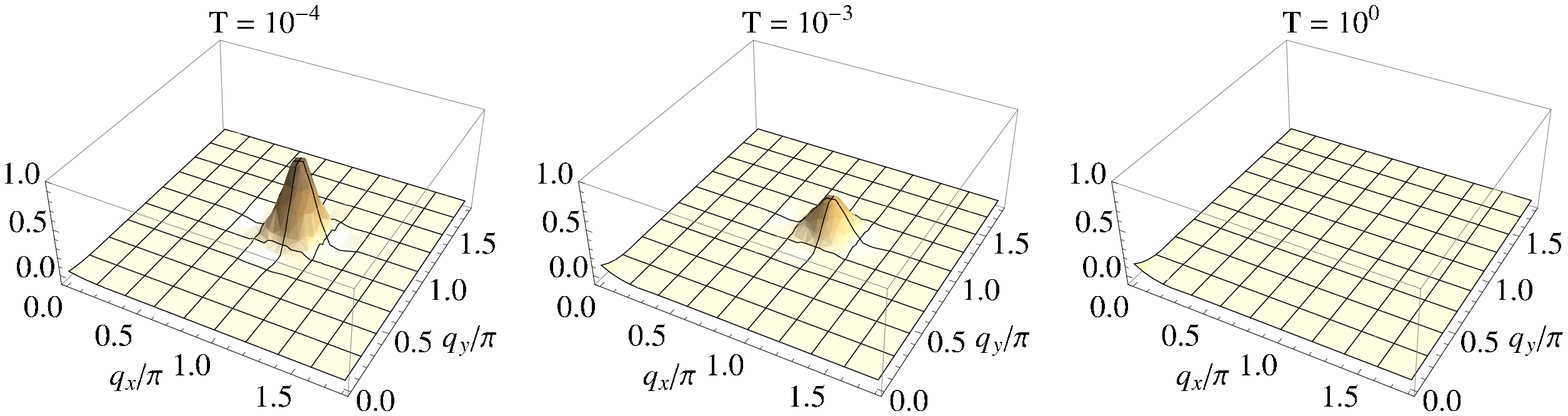}
\caption{The orientational order parameter $O({\bf q})$ as a function
  of the two-dimensional wave vector ${\bf q}$ at three temperatures
  across the melting transition ($\kappa l=10$, $A=1$). At the lowest
  temperature (AF phase) the peak $O({\bf q_0}=(\pi,\pi))$ reflects
  the modulation of the angular orientation of the dimers. Upon
  raising the temperature, this peak gradually decays and at high
  temperature (modulated liquid phase), where the dimers are
  uncorrelated, a peak at ${\bf q}={\bf 0}$ emerges.}
\label{fig4}
\end{figure}

\begin{figure}[ht]
\includegraphics[width=0.7\textwidth]{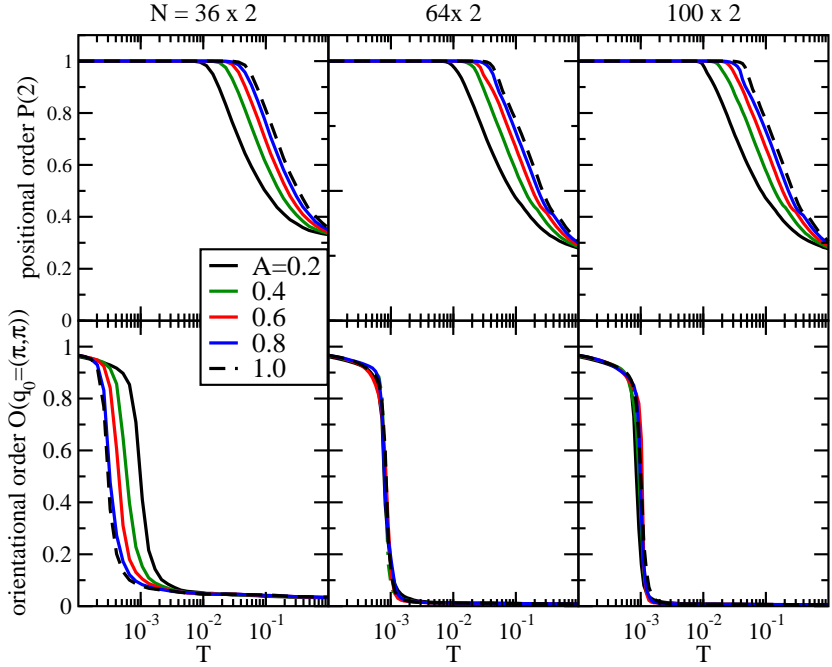}
\caption{Melting transitions of AF phase for $\kappa l=10$ and various
  trap strengths $A=0.2...1$ and three system sizes. {\bf Upper row:}
  the positional order parameter $P(2)$ as a function of $T$. {\bf
    Lower row:} the temperature dependence of the orientational order
  parameter $O({\bf q_0}=(\pi,\pi))$ defined in Eq. (\ref{oq}).}
\label{fig5}
\end{figure}

\begin{figure}[ht]
\includegraphics[width=0.7\textwidth]{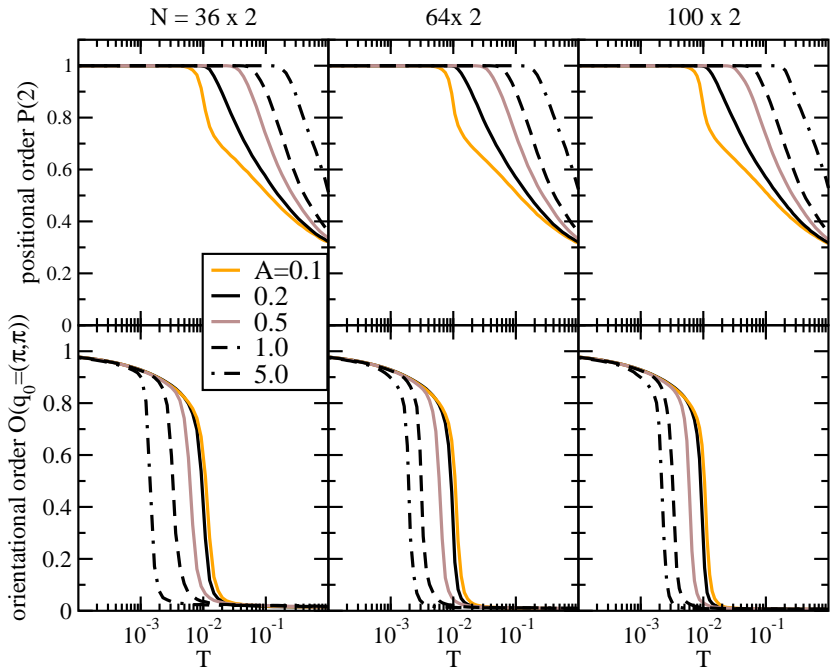}
\caption{Same as Fig.\ref{fig5} for $\kappa l=6$ and trap strengths
  $A$ from 0.1 to 5. }
\label{fig6}
\end{figure}

\begin{figure}[ht]
\vspace{1mm}
\includegraphics[width=0.75\textwidth]{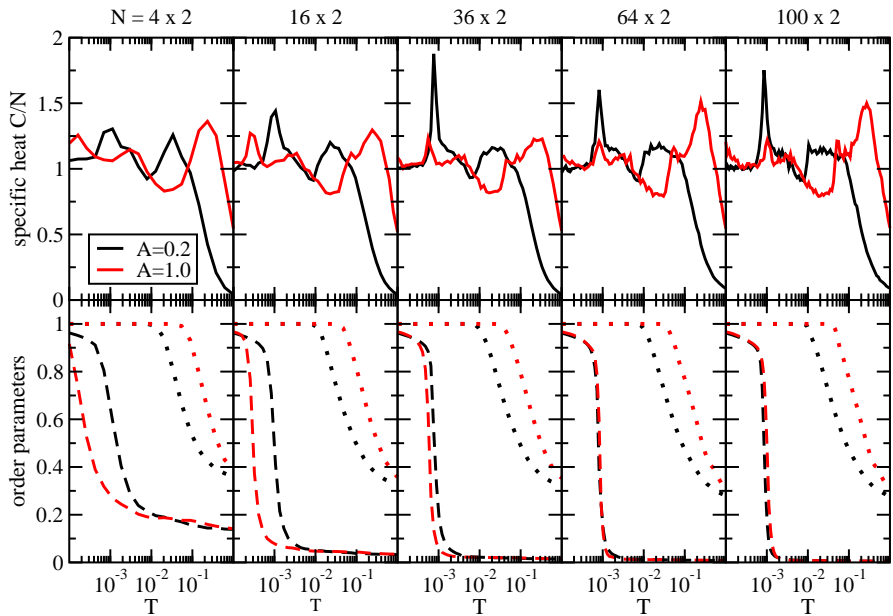}
\caption{{\bf Upper row:} Temperature dependence and finite size
  scaling of the specific heat per colloid $C/N$ at $\kappa l=10$ and
  two values of $A$ {\bf Lower row:} The corresponding variation of
  the orientational $O({\bf q_0})$ (dashed lines) and positional
  $P(2)$ (dotted lines) order parameters.}
\label{fig7}
\end{figure}

In Fig.\ref{fig3} we show typical snapshot configurations at different
temperatures across the melting transition. The right-most column
shows the energy and moments of positional and angular distributions
as a function of temperature. In the snapshots, one can observe AF
configurations with long-range positional and orientational order,
positionally ordered (exactly 2 colloids per trap) but orientationally
disordered states and liquid-like disordered configurations. In order
to quantify the positional order, we show normalized distributions of
the trap occupancy $P(N_t)$, which is narrow and peaked at $N_t=2$ at
low enough temperatures, while it broadens around the average $N_t=2$
at higher temperatures. From these distributions it seems natural to
extract a $T$-dependent order parameter as $P(2)$, measuring the ratio
of traps in the system with exactly 2 colloids. Any deviations from
the saturated value $P(2)=1$ signals the presence of positional
melting.  Smooth decrease of $P(2)$ with increasing $T$ is a sign of a
crossover transition rather than a phase transition; this is
additionally supported by the lack of symmetry breaking at the
transition. In this respect, a crossover transition temperature
$T_{c1}$ cannot be rigorously defined, yet we decide to mark with
$T_{c1}$ the beginning of crossover at the point where $P(2)$ starts
to deviate from 1. At that temperature on average the $N_{\rm tr}-2$
traps still host exactly 2 colloids, while two of the traps exchange
one colloid to host 1 and 3 colloids, respectively and the threshold
value of the order parameter is $P(2)=1-2/N_{\rm tr}$. From the
results at $\kappa l=10$ and $A=0.6$ we estimate $T_{c1}\approx
2\times 10^{-2}$.  The so-defined quantity is showing negligible
finite size effects.
At low temperatures the colloidal dimers are orientationally
correlated. This is indirectly demonstrated in the bottom row of Fig.
\ref{fig3} where the angle distributions $P(\phi)$ of dimers at
various temperatures \cite{inertia} are depicted: at the lowest
temperature two narrow peaks of equal height imply that half of the
dimers lie horizontally ($\phi=0$) and the other half vertically
($\phi=\pi/2$), as expected for the AF phase. To characterize the
above distributions with a meaningful scalar measure, we define 
the spread $\sigma_p$ of the distribution $P(\phi)$:
$\sigma_P^2=1/N_{\rm bin} \sum_i (P_i-1)^2$, where 
$N_{\rm bin}$ is the number of bins used to sample the
variable $\phi$. In order to have a quantity bounded by 1, we define
the spreading parameter $\sigma_P/\sigma^0_P$, where $\sigma^0_P =
\sqrt{N_{\rm bin}/2 -1}$ is the zero temperature limit of $\sigma_P$
in the AF phase. At large $T$ the value of $\sigma_P/\sigma^0_P$ is
zero indicating orientationally disordered configuration and it starts
increasing upon lowering the temperature below $T_{c1}$. However, the
spreading parameter is not a suitable order parameter for studying the
details of the orientational melting transition, since it only deals
with one-dimer statistics and therefore does not capture dimer-dimer
angular correlations. To properly account for the angular correlations
in dimeric coverings, we define a new order parameter analogue to the
Fourier transform of the spin-spin correlator \cite{oq}. We start from
the averaged static correlation function that measures average
angle-angle correlation between two traps separated by
${\mbox{\boldmath$\delta$}}$
\begin{eqnarray} 
   O({\mbox{\boldmath$\delta$}})=\frac{1}{N_{\rm tr}} \sum_{i=1}^{N_{\rm tr}}
   \left\langle\big| \cos(\phi_{i+{\mbox{\boldmath$\delta$}}} - \phi_i)\big|\right\rangle
   \label{cor}
\end{eqnarray} 
where $N_{\rm tr}$ denotes the number of traps in a system and
$\phi_i$ measures orientation of the dimer in $i$-th trap and $\langle...\rangle$
denotes thermal averaging. Since dimers are not directed and therefore
$\phi$ and $\phi\pm\pi$ indicate the same orientation, we use the
absolute value in Eq. (\ref{cor}). The function
$O({\mbox{\boldmath$\delta$}})$ is bounded on $[0,1]$; in order to
compute its Fourier transform, we first expand its range to $[-1,1]$
by transforming it: $O({\mbox{\boldmath$\delta$}}) \to 2
O({\mbox{\boldmath$\delta$}})-1$. The Fourier transform then reads:
\begin{eqnarray} 
O({\bf q})=\frac{1}{N_{\rm tr}}
\sum_{\mbox{\boldmath$\delta$}}(2
  O({\mbox{\boldmath$\delta$}})-1) {\rm e}^{{\rm i} {\bf
        q}\cdot{\mbox{\boldmath$\delta$}}}
\label{oq} 
\end{eqnarray} 
${\bf q}$ being a 2D ordering wave vector. The temperature
variation of $O({\bf q})$ across the melting transition from AF to
modulated liquid is depicted in Fig.\ref{fig4}. In an ideally ordered
AF phase at $T=0$, characterized by modulation ${\bf q}={\bf q_0}$,
the order parameter saturates, $O({\bf q_0}=(\pi,\pi))=1$. At high
temperatures, on the contrary, any pair of traps is uncorrelated, which
gives a vanishing $O({\bf q})$ for all non-zero wavevectors: $O({\bf
  q}\ne {\bf 0})\propto 1/N_{\rm tr}$ \cite{oq2}.
In Fig.\ref{fig5} we compare the orientational and positional
order-disorder transitions at $\kappa l=10$ for several confinement
strengths $A$ at three system sizes. In the upper row we plot the
maximum value of the trap occupancy distribution $P(2)$ as an order
parameter for the positional order. We see that the melting transition
is gradual, depends on the confinement strength and shows very little
size dependence. In the lower row we show the temperature dependence
of the orientational order parameter $O({\bf q_0})$ (Eq.\ref{oq})
using the antiferro ordering wave vector ${\bf q_0}=(\pi,\pi)$. It
varies from 1 (AF phase) at low temperatures to 0 (orientational
disorder) at high $T$. In contrast to the positional melting, the
transition is now sharp and it grows sharper with increasing system
size. This behaviour is a signature of a second order phase
transition with diverging correlation length. We also note that at the
transition the system reduces translational symmetry with respect to
the periodicity of the underlying trap potential. In Fig.\ref{fig6},
the same quantities are plotted for $\kappa l=6$. Very similar results
have been obtained at all other parameter values $\kappa l$ and $A$
marked on Fig.\ref{fig2} (results not shown). We therefore conclude
that there are --in agreement with previous results \cite{RO2002}--
two transitions going on in the system upon cooling from the liquid
phase: at $T_{c1}\approx 10^{-2}$ a gradual crossover ordering
transition to the positionally ordered structures sets in, 
and at $T_{c2}\approx 10^{-3}$ a second order
phase transition from orientationally disordered to ordered state.
In order to further strengthen the evidence for a second order phase
transition, we have also calculated the specific heat as a function of
temperature. The specific heat of a finite canonical system is given
by the energy fluctuations:
\begin{eqnarray}
   C=\frac{\langle E^2 \rangle-\langle E \rangle^2}{T^2}\;. 
\label{spec}
\end{eqnarray}
We have evaluated this expression and observed a narrow peak in $C$ at
$T_{c2}$ showing the expected system size dependence for a second
order phase transition (in an infinite system the specific heat
would diverge at $T_{c2}$). In contrast, there is a broad
size-independent peak positioned at $T_{c1}$. In Fig. \ref{fig7}, we
present the specific heat curves as a function of temperature and
system size at two extreme values of $A$ and demonstrate that a
maximum increase of both order parameters coincides with peak
positions in specific heat function.
An interesting observation is that the position of the crossover
transition $T_{c1}$ shifts towards higher temperatures with
increasing
 trap strength $A$, while the position of the phase
transition $T_{c2}$ shows a weaker dependence on $A$. From analysing
the finite size scaling (see the Supplementary material) of the two
transition temperatures we conclude that both, $T_{c1}$ and $T_{c2}$
are finite in the thermodynamic limit \cite{critical_exp}.
\section{Discussion}
\begin{figure}[ht]
\includegraphics[width=0.75\textwidth]{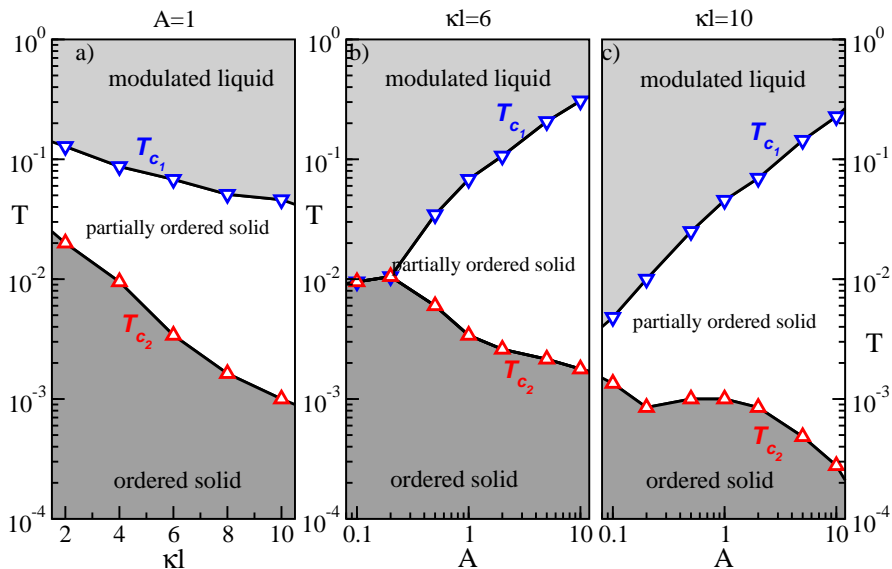}
\caption{Phase behaviour of dimers on the square lattice as calculated
  in the largest system with $N=200$ colloids: {\bf a)} Transition
  temperatures versus $\kappa l$ at fixed confinement strength $A=1$;
  {\bf b)}Transition temperatures versus $A$ at fixed screening
  $\kappa l=6$ and {\bf c)} $\kappa l=10$.}
\label{fig9}
\end{figure}

To summarize, we present (Fig.\ref{fig9}) phase diagrams showing both
transition temperatures versus $\kappa l$ at fixed confinement
strength and versus $A$ at fixed $\kappa l$. The crossover transition
$T_{c1}$ has been somewhat arbitrary defined as the temperature at
which the positional melting starts and the value of the order
parameter $P(2)$ drops from 1 to $1-2/N_{\rm tr}$. On the other hand,
$T_{c2}$ was extracted as the point of the steepest decrease of the
orientational order parameter $O({\bf q_0})$ exactly matching the
position of the diverging peak in the specific heat $C$
(Fig.\ref{fig7}). With increasing screening strength $\kappa l$, both
critical temperatures decrease (see left panel in
Fig. \ref{fig9}). This is due to weaker inter-trap interactions that
directly control orientational ordering ($T_{c2}$) and also, in
addition to trapping forces, average distances between the dimers,
which affects positional ordering ($T_{c1}$). Increasing $A$ has a
similar effect on $T_{c2}$ (see middle and right panels in
Fig. \ref{fig9}) since larger $A$ means smaller dimers and hence
weaker inter-trap forces. On the contrary, positional order is
of course stronger at larger $A$, therefore higher temperatures are
needed to destroy it. At large enough values of $\kappa l$ and $A$ we
observe the two-stage melting with the phase transition from ordered
solid to partially ordered solid (without orientational order) and the
crossover melting of the positional order from partially ordered solid
to modulated liquid. Interestingly, at longer screening lengths
(smaller $\kappa l$) and weaker confinement the crossover transition
disappears and the ordered solid melts directly into a modulated
liquid via a second order phase transition. This can also be observed
in Fig. \ref{fig6} (first row, black line) where the $P(2)$ curve
exhibits a sudden change in the slope, precisely at $T_{c2}$.  
The phase diagram reported here is qualitatively consistent with the
result of Langevin simulations reported in \cite{RO2002}. The
screening lengths considered in our work ($\kappa l=6,10$) are much
shorter than in \cite{RO2002} ($\kappa l=2$), therefore the phase
diagrams in Fig.\ref{fig9} are shifted to lower values of the
confinement $A$ as compared to Fig.4 in \cite{RO2002}. In
\cite{Sarlah} spin models were constructed to analytically describe
orientational ordering of colloidal molecules on periodic
lattices. Remarkably, their analytically obtained orientational
melting transition for trimers on triangular lattice coincides with
the corresponding transition for dimers on square lattices in
\cite{RO2002}. Crystaline structures of dimers on triangular lattice
were also examined in \cite{Sarlah} showing interesting transitions
between herringbone, ferromagnetic, ``Japanese~6~in~1'', paramagnetic and
antiferromagnetic ordering. However, several approximations had to be
made in order to reduce the colloidal system to a Potts-like model:
they assumed rigid dimers placed on the lattice points interacting
with nearest neighbours only and restricted to discrete orientations
compatible with the lattice symmetry. In \cite{ElShawish} we have
already discussed that the discrete angle approximation does not apply
to the experimental system of colloids trapped by laser
tweezers. Further to that, while the nearest neighbour approximation
seems well justified for triangular lattices, it is qualitatively
wrong on square substrates \cite{ElShawish}, therefore it seems
difficult to extend the conclusions of \cite{Sarlah} to colloidal
systems and to square-like patterned substrates.

\section{Acknowledgements}

This work was supported by the Slovenian Research Agency through the
Grant P1-0055, by European Research Council through the Advanced
Research Grant COLSTRUCTION (RG52356), and by the 7th Framework
Programme through the ITN network COMPLOIDS (RG234810).


\begin{thebibliography}{99}
\bibitem{inertia}
  A $n$-mer orientation is defined as an angle $\phi$ between the
  horizontal axis and $n$-mers's principal axis with the smallest
  principal moment. Principal axes and principal moments are obtained
  by diagonalizing the $2\times 2$ area moment of inertia tensor
  $J_{ij}=\int(r^2 \delta_{ij}-r_i r_j) dA$ where $i,j=x,y$ and
  distance $r_i$ is measured relatively to the center of mass. Since
  colloids are viewed as point-like objects, the area integral in
  $J_{ij}$ is further reduced to a sum over all colloid positions
  within a given $n$-mer. This definition of orientation is robust for
  all $n>1$ and all shapes.
\end{thebibliography}
\end{document}